\documentclass[12pt]{iopart}
\usepackage{iopams}
\expandafter\let\csname equation*\endcsname\relax

\expandafter\let\csname endequation*\endcsname\relax
\usepackage{amsmath}
\usepackage{amssymb}

\usepackage{empheq}
\usepackage{dsfont}
\usepackage{graphicx}
\usepackage{hyperref}
\usepackage{stackrel}
\usepackage{color,epsfig}
\usepackage{comment}

\usepackage{colortbl}

\input{epsf}

\numberwithin{equation}{section}

\newcommand{\BS}{\boldsymbol \sigma}
\newcommand{\Bgrad}{\boldsymbol \nabla}

\def\p{\mathbf{p}}
\newcommand{\bb}{\begin{equation}}
\newcommand{\ee}{\end{equation}}
\newcommand{\eqb}{\begin{eqnarray}}
\newcommand{\eqf}{\end{eqnarray}}

\def \ee{\end{equation}}
\def \be{\begin{equation}}
\def \eea{\end{eqnarray}}
\def \bea{\begin{eqnarray}}

\begin{document}

\title[Optical Transparency in an effective model for Graphene]{Optical Transparency in an effective model for Graphene}

\author{Horacio Falomir}
\address{IFLP, CONICET - Departamento de F\'{\i}sica, Fac.\ de Ciencias Exactas de la UNLP, C.C. 67, (1900) La Plata, Argentina.}
\ead{falomir@fisica.unlp.edu.ar}
\author{Marcelo Loewe}
\address{Instituto de F\'isica, Pontificia Universidad Cat\'olica de Chile, \\
Casilla 306, Santiago, Chile.}
\address{Centre for Theoretical and Mathematical Physics, University of Cape Town, Rondebosch 770, South Africa.}
\address{Centro Cient\'ifico Tecnol\'ogico de Valpara\'iso, CCTVAL, Universidad T\'ecnica Federico Santa Mar\'ia, Casilla 110-V, Valpara\'iso, Chile.}
\ead{mloewe@fis.puc.cl}
\author{Enrique Mu\~noz}
\address{Instituto de F\'isica, Pontificia Universidad Cat\'olica de Chile, \\
Casilla 306, Santiago, Chile.}
\ead{munozt@fis.puc.cl}
\author{Alfredo Raya}
\address{Instituto de F\'isica y Matem\'aticas, Universidad Michoacana de San Nicol\'as de Hidalgo. Edificio C-3, Ciudad Universitaria. Francisco J. M\'ujica s/n. Col. Fel\'icitas del R\'io. C.P. 58040, Morelia, Michoac\'an, Mexico.}

\begin{abstract}
Motivated by experiments confirming that the optical transparency of graphene is defined through the fine structure constant and that it could be fully explained within the relativistic Dirac fermions in 2D
picture, in this article we
investigate how this property is affected by next-to-nearest
neighbor coupling in the low-energy
continuum description of graphene. A detailed calculation within the linear response regime allows us to conclude that, somewhat surprisingly,
the zero-frequency limit of the optical conductivity that determines the transparency remains robust up to this correction.
\end{abstract}



\maketitle
\section{Introduction}
Graphene is a two-dimensional allotrope of carbon, arranged as a honeycomb lattice with a $C_{3v}\otimes Z_2$ symmetry \cite{Wallace_47}
that determines its remarkable physical properties \cite{Munoz_16,Peres_10,Goerbig_11,Novoselov_05}. In particular, the electronic spectrum arising from an atomistic tight-binding
model displays two non-equivalent points $K_{+},\,K_{-}$ where the conduction and valence bands touch, and in whose vicinity
the dispersion relation is approximately linear. This leads to an effective, low-energy continuum model where the electronic properties
of the material are well captured by those of relativistic Dirac fermions in 2D. Among the plethora of physical consequences of this fact that have been already
predicted and measured \cite{Munoz_16,Peres_10,Goerbig_11,Novoselov_05,Vozmediano_10,DasSarma_11,Novoselov_04}, we noticed an interesting experiment that measures the optical transparency of single and few-layer graphene~\cite{Nair}.
The transparency is a physical property that is determined by the optical conductivity, i.e. the linear response to an electromagnetic field, in the zero-frequency limit.
 A variety of experiments confirm~\cite{Nair,WASSEI201052,Ma2013,Mak24082010,shou} that the measured transmittance is indeed compatible with the effective single-particle model of relativistic Dirac fermions in
graphene.
A number of different theoretical works have exploited this fact to calculate the light absorption rate in graphene from a ``relativistic'' quantum electrodynamics perspective~\cite{FV-2012,mariel,FV-2016,FV-2011B,FV-2011,Fial-2011,david,saul,Merthe}.
An interesting question that remains open is up to what extent this effective model is valid in the representation of this optical property, since
it arises from a tight-binding microscopic atomistic model that involves only the nearest neighbor hopping. In this article, we decided to explore what is the
contribution to the optical conductivity arising from the next-to-nearest neighbors coupling in the atomistic Hamiltonian,
included as a quadratic correction to the kinetic energy operator within the continuum effective model for graphene. Such a model has been considered in Ref.~\cite{GNAQ} to fully account for the Anomalous Integer Quantum Hall Effect in this material and the underlying wave equation is referred to in literature as Second Order  Dirac Equation~\cite{second}. For our purposes, let us recall that within the
 linear response theory, general Kubo relations allow to express the transport coefficients in terms of retarded correlators~\cite{Wen}, that for a pair of observables $\hat{O}_1$, $\hat{O}_2$ are defined
by ($\zeta = \pm$ for Bosons and Fermions, respectively)
\begin{eqnarray}
C_{O_1,O_2}^{R}(t-t') &=& -i\theta(t - t')\langle [ \hat{O}_1(t),\hat{O}_2(t')]_{-\zeta}  \rangle\nonumber \\
&=& -i\theta(t - t')\langle \hat{O}_1(t)\hat{O}_2(t')  \rangle -i\theta(t - t')\zeta\langle \hat{O}_2(t')\hat{O}_1(t)  \rangle.
\label{eq_C1}
\end{eqnarray}
These retarded correlators differ from the usual time-ordered ones that, by construction, are obtained via functional differentiation of the standard generating functional
constructed form a path-integral formulation in quantum field theory. This rather technical inconvenience can be overcome by connecting the different propagators using a Lehmann representation, or alternatively to work in the Matsubara formalism at finite temperature and use analytic continuation a posteriori~\cite{Wen}. There is however a third, and
more direct alternative, which is to express the generating functional in the contour time path (CTP),  also known as Keldysh formalism in the condensed matter literature~\cite{Rammer,Stefanucci}. In this work, we choose the CTP formalism to explicitly calculate the polarization tensor as a retarded correlator of the current operators, which provides the correct definition
of the optical conductivity within linear response theory.
With these ideas in mind, we have organized the remaining of this article as follows: In Sect.~\ref{Ec-Mov}, we present the details of the model. In Sect.~\ref{CTP} we present the Keldysh formalism to calculate the current-current correlator and in Sect.~\ref{vacpol} we obtain the optical conductivity from the vacuum polarization tensor. We discuss our findings in Sect.~\ref{conclusions}. Some calculational details are presented in an Appendix.

\section{Lagrangian, conserved current and generating functional}\label{Ec-Mov}

\begin{figure}[tbp]
\centering
\epsfig{file=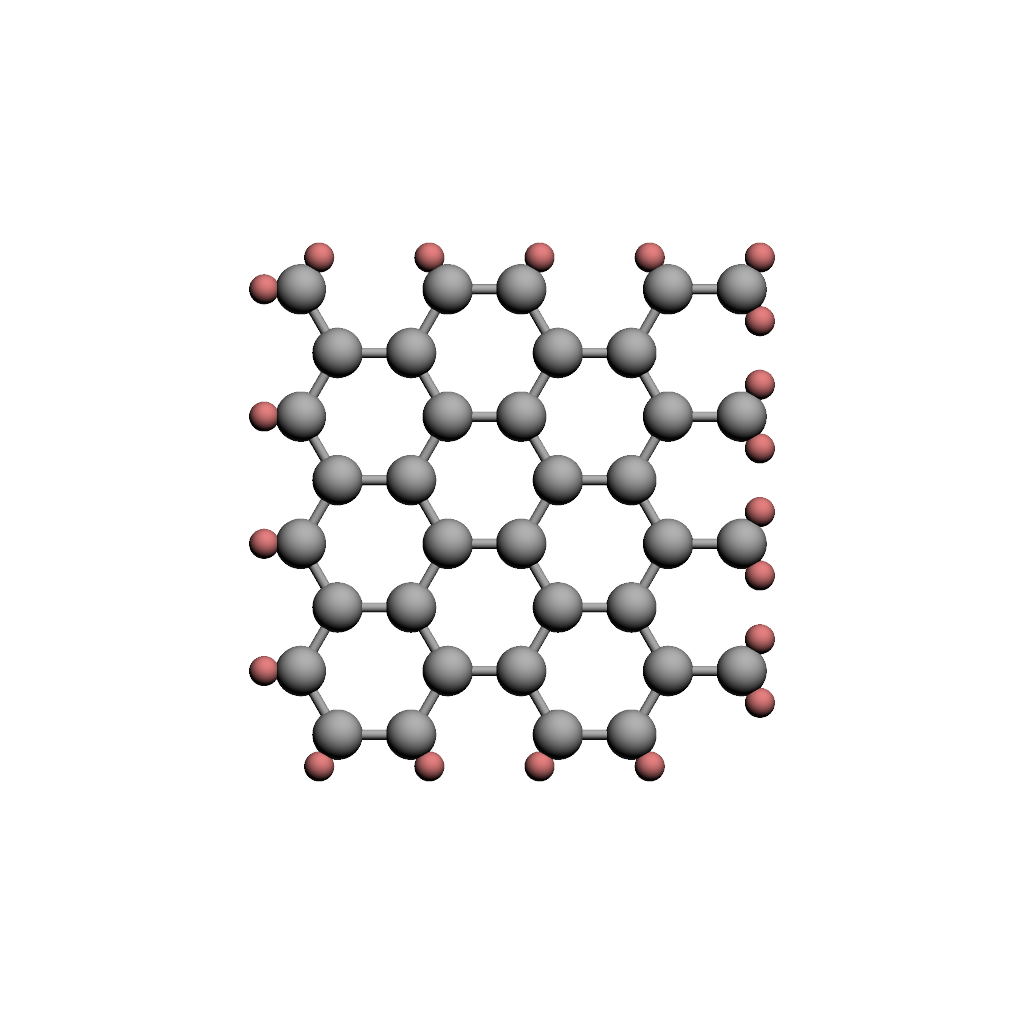,width=0.5\columnwidth}
\caption{(Color online) Sketch of the crystal structure of graphene. The honeycomb array is described in terms of two overlapping triangular sublattices. 
}
\label{fig1a}
\end{figure}

Graphene consist in one atom thick membrane of tightly packed carbon atoms in a honeycomb array.  Its crystal structure, sketched in Fig.~\ref{fig1a}, is described in terms of two overlapping triangular (Bravais) sublattices so that for a given atom belonging to any of these sublattices, its nearest neighbors belong to the second sublattice, the next-to-nearest neighbors to the original sublattice and so on. The band structure at the next-to-nearest approximation is of the form
\begin{equation}
E_\pm(\mathbf{k})=\pm t\sqrt{f(\mathbf{k})}-t'[f(\mathbf{k})-3],
\end{equation}
where $t$ and $t'$ are the nearest and next-to-nearest hopping parameters and
\begin{equation}
f(\mathbf{k})=3+4\cos\left( \frac{3k_x a}{2}\right)\cos\left( \frac{\sqrt{3}k_y a}{2}\right)+2\cos(\sqrt{3}k_ya)\;,
\end{equation}
where $a\simeq 1.42\AA$ is the interatomic distance. The points $K_+$ and $K_-$ at which $f(K_\pm)=0$ define the so-called Dirac points. Around $K_+$,
\begin{equation}
E_\pm(\mathbf{k}+K_+)=\pm t\left[\frac{3}{2}a |\mathbf{k}|-\frac{3}{8}a^2 \mathbf{k}^2 \sin(3\vartheta) \right]+t'\left[-\frac{9}{4}a^3\mathbf{k}^2+3 \right]+{\cal O}(|\mathbf{k}|^3)\;,\label{nnn}
\end{equation}
with $\tan(\vartheta)=k_y/k_x$. Around $K_-$ one merely has to replace $\vartheta\to -\vartheta$ in Eq.~(\ref{nnn}). The isotropic portion of this model was first considered in Ref.~\cite{GNAQ} as a natural framework to explain the Anomalous Integer Quantum Hall Effect in graphene. The anisotropic term in that work was treated perturbatively and shown not to contribute to the energy spectrum at first order. In the presence of electromagnetic interactions, the model is described by
the Lagrangian~\cite{GNAQ}
\begin{eqnarray}
      \mathcal{L}&:=&\frac{i}{2}\left[\psi^\dagger \, \partial_t \psi - \partial_t\psi^\dagger\, \psi\right]+
      \psi^\dagger e A_0  \psi \nonumber\\
  &&-\frac{1}{2m} \left\{ \left[\left(\mathbf{p}-e\mathbf{A}+\theta\BS\right)\psi\right]^\dagger \cdot \left[\left(\mathbf{p}-e\mathbf{A}+\theta\BS\right)\psi\right]-2\theta^2\psi^\dagger\psi\right\}\nonumber\\
   &=& \frac{i}{2}\left[\psi^\dagger \, \partial_t \psi - \partial_t\psi^\dagger\, \psi\right]
   -\frac{1}{2m} \left\{\Bgrad\psi^\dagger \cdot \Bgrad \psi +
   i \Bgrad\psi^\dagger \cdot  \left(-e\mathbf{A}+\theta\BS\right)\psi -\right.\nonumber\\
   &&\left. -i \psi^\dagger  \left(-e\mathbf{A}+\theta\BS\right) \cdot\Bgrad\psi+
   \psi^\dagger  \left[\left(-e\mathbf{A}+\theta\BS\right)^2- 2 \theta^2 \right]\psi
   \right\}
\,,\label{L1}
\end{eqnarray}
where $\theta=m v_F$. Here, $\psi^\dagger$ and $\psi$ are regarded as independent fields whose equations of motion are derived from the variation of the action with respect to these fields, namely,
\begin{eqnarray}
        \frac{\partial \mathcal{L}}{\partial \psi^\dagger}-\partial_t \left( \frac{\partial\mathcal{L}}{\partial \left(\partial_t \psi^\dagger\right)}\right) -
    \Bgrad\cdot \left( \frac{\partial\mathcal{L}}{\partial \left(\Bgrad \psi^\dagger\right)}\right)\nonumber\\
   &&\hspace{-5cm}   =i \partial_t \psi- \frac{1}{2m} \left[\left(\mathbf{p}-e\mathbf{A}+\theta\BS\right)^2
      -2\theta^2\right] \psi =0\,,\label{L2}
\end{eqnarray}
and similarly for $\psi$.

\medskip

The Lagrangian in Eq.(\ref{L1}) remains invariant against the  local change in the dynamical variables and the external electromagnetic field
\begin{equation}\label{L4}
    \begin{array}{c} \displaystyle
    \psi(x)\rightarrow e^{i e \alpha(x)}\psi(x) \quad \Rightarrow \quad  \delta \psi(x) = i e \alpha(x) \psi(x)\,,
      \\ \\ \displaystyle
      \psi^\dagger(x)\rightarrow \psi^\dagger(x) e^{-i e \alpha(x)} \quad \Rightarrow \quad  \delta \psi^\dagger(x)
       = -i e \alpha(x) \psi^\dagger(x)\,,
       \\ \\ \displaystyle
        A_\mu(x) \rightarrow A_\mu(x) + \partial_\mu \alpha(x)\,,
    \end{array}
\end{equation}
that is, it  has a $U(1)$ gauge symmetry.  N{\oe}ther's Theorem leads to the existence of the locally conserved current
\begin{equation}\label{L5}
    \alpha j^\mu := - \delta \psi^\dagger \left( \frac{\partial\mathcal{L}}{\partial \left(\partial_\mu \psi^\dagger\right)}\right)
    - \left( \frac{\partial\mathcal{L}}{\partial \left(\partial_\mu \psi\right)}\right) \delta\psi\,.
\end{equation}
The corresponding charge density is
\begin{equation}\label{L6}
    j^0 = e \, \psi^\dagger \psi 
\end{equation}
and the current density
\begin{equation}\label{L7}
    {j^k}=\frac{e}{2m}\left\{
    i \left(\partial_k \psi^\dagger\,  \psi - \psi^\dagger \, \partial_k\psi \right)
    + 2 \psi^\dagger\left(-e {A_k}+\theta\sigma_k\right)\psi\right\}\,.
\end{equation}
It is straightforward to verify, from  the equations of motion, that $j^{\mu}$ is conserved,
\begin{equation}\label{L8}
    \partial_\mu j^\mu=\partial_t j^0 - \Bgrad \cdot \mathbf{j}=0\,.
\end{equation}
Notice also that we can write
\begin{equation}\label{JderivGamma}
    j^\mu(x)= \frac{\delta}{\delta A_\mu(x)} \int \mathcal{L}(y) \,  d^3 y\,.
\end{equation}
With these ingredients, we can formulate the corresponding current-current correlator.

\section{Generating functional in the Contour Time Path.}~\label{CTP}
\begin{figure}[tbp]
\centering
\epsfig{file=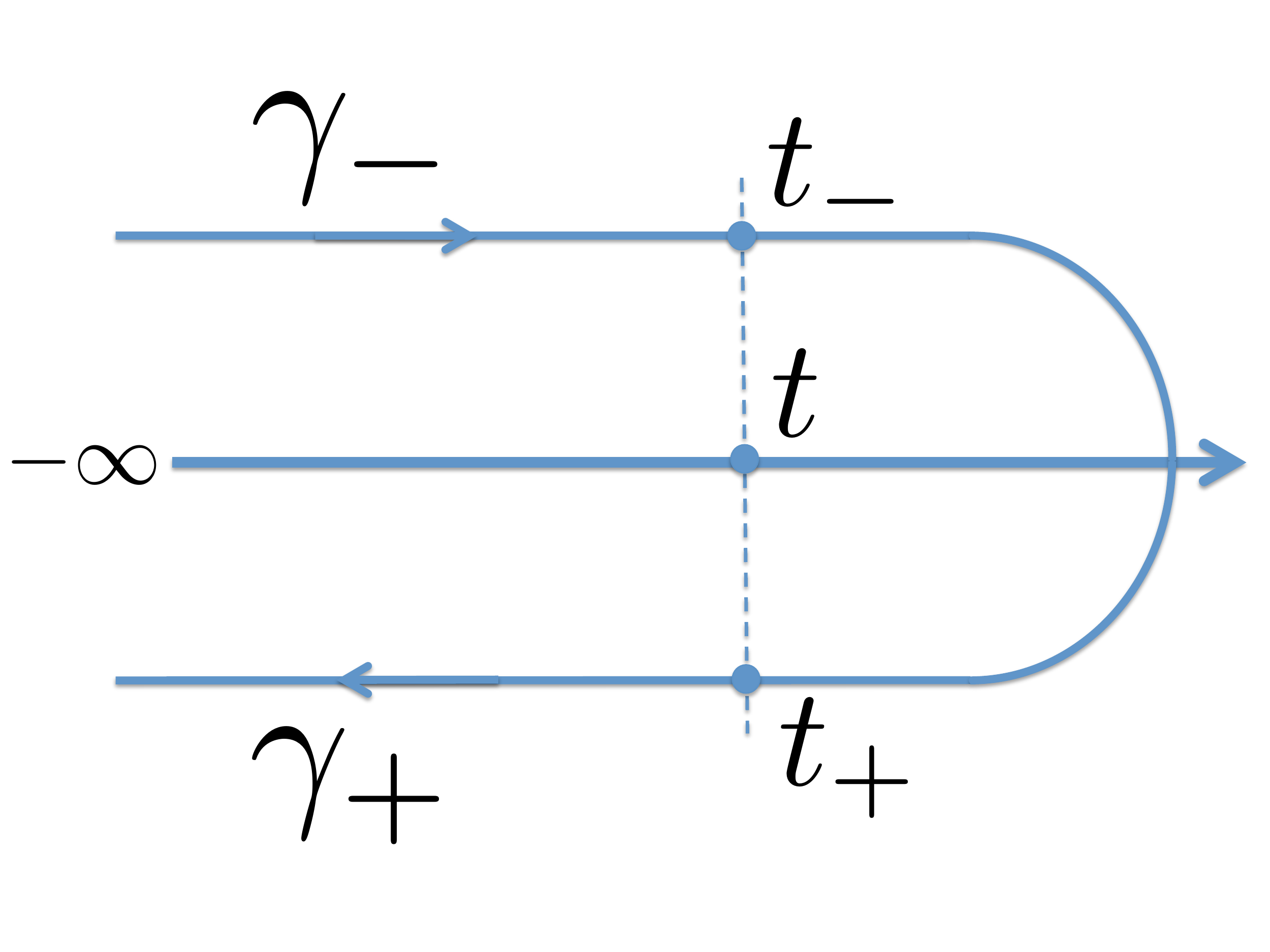,width=0.3\columnwidth}
\caption{(Color online) The contour $\gamma = \gamma_{+}\oplus\gamma_{-}$ is depicted in the figure. The double folding of the time axis
is displayed, by showing that always two points $t_{-}$ and $t_{+}$, located in the time ordered $t_{-}\in\gamma_{-}$ and and anti-time ordered $t_{+}\in\gamma_{+}$ branches
of the contour correspond to the same chronological time instant $t$.}
\label{fig1}
\end{figure}

We seek to calculate the polarization tensor, defined as a retarded current-current correlator that, in linear response, determines the optical conductivity. For that
purpose, we choose to represent the field-theory described in the previous section on the Contour Time Path (CTP)~\cite{Rammer,Stefanucci}. Let us define the contour $\gamma = \gamma_{-}\oplus\gamma_{+}$, where $\gamma_{-}$ represents the time-ordered branch
while $\gamma_{+}$ the anti-time-ordered branch, as depicted in Fig.\ref{fig1}. Therefore, we define a contour evolution
parameter $\tau\in\gamma$, such that
\begin{eqnarray}
\tau = \left\{\begin{array}{cc}
t_{-},&\tau\in\gamma_{-}\;,\\
t_{+},&\tau\in\gamma_{+}\;.
\end{array}\right.
\end{eqnarray}
Also notice that,
as depicted in Fig.\ref{fig1}, both $t_{+}$ and $t_{-}$ have a unique correspondence to a given chronological instant of time $t\in {\rm I\!R}$.
Correspondingly, for operators and fields defined with their time arguments along the CTP, we have the definitions
\begin{eqnarray}
\psi(\mathbf{x},\tau) = \left\{\begin{array}{cc}
\psi(\mathbf{x},t_{-}) \equiv \psi_{-}(\mathbf{x},t),&\tau\in\gamma_{-}\;,\\
\psi(\mathbf{x},t_{+}) \equiv \psi_{+}(\mathbf{x},t),&\tau\in\gamma_{+}\;.
\end{array}
\right.
\end{eqnarray}
Then, the generating functional of (current) Green's functions of this two-dimensional system, defined on the CTP reads
\begin{equation}\label{eq_CTP_gen}
    Z_{\gamma}[A]=e^{i \Gamma_{\gamma}[A]}:=\int \mathcal{D} \psi^\dagger(\mathbf{x},\tau) \mathcal{D}\psi(\mathbf{x},\tau) \, e^{\displaystyle
    i \int_{\gamma} d\tau \int d^2 \mathbf{x}\, \mathcal{L}[\psi^\dagger(\mathbf{x},\tau),\psi(\mathbf{x},\tau)]}\,,
\end{equation}
where $\Gamma_{\gamma}[A]$ is the effective contribution to the action for the electromagnetic field.
The path-integral on the CTP induces by construction the contour-ordering between the fields, defined by the operation $\mathcal{T}$ between two operators $\hat{O}_1(\tau)$ and $\hat{O}_2(\tau)$ in the Heisenberg picture ($\zeta = \pm$ for Bosons/Fermions, respectively)
\begin{eqnarray}
\langle \mathcal{T} \hat{O}_1(\tau_1)\hat{O}_2(\tau_2) \rangle &=& \theta(\tau_1 - \tau_2)\langle \hat{O}_1(\tau_1)\hat{O}_2(\tau_2) \rangle \nonumber\\
&& + \zeta  \theta(\tau_2 - \tau_1)\langle \hat{O}_2(\tau_2)\hat{O}_1(\tau_1)\rangle.
\end{eqnarray}
Here, we have defined the contour Heaviside function as
\begin{eqnarray}
\theta(\tau_1 - \tau_2) = \left\{
\begin{array}{cc}
1,&\tau_1 >_{c} \tau_2\;,\\
0,&\tau_2  >_{c} \tau_1\;,
\end{array}
\right.
\end{eqnarray}
with the symbol $>_{c}$ indicating the relation ``later than in the contour''.
In general physical situations where the sources and external fields do not break time-reversal invariance, $\psi_{-}(x,t) = \psi_{+}(x,t)$,
and the CTP becomes just a useful trick to express at once all the different correlators. Consider for instance the contour-ordered correlator
between two fields,
\begin{eqnarray}
\Delta(\mathbf{x}_1,\tau_1;\mathbf{x}_2,\tau_2) &\equiv& -i\langle \mathcal{T} \psi(\mathbf{x}_1,\tau_1) \psi^{\dagger}(\mathbf{x}_2,\tau_2) \rangle
\nonumber\\
&=& \theta(\tau_1 - \tau_2) (-i)\langle  \psi(\mathbf{x}_1,\tau_1) \psi^{\dagger}(\mathbf{x}_2,\tau_2) \rangle\nonumber\\
&& + \zeta \theta(\tau_2 - \tau_1) (-i)\langle \psi^{\dagger}(\mathbf{x}_2,\tau_2) \psi(\mathbf{x}_1,\tau_1)  \rangle\;.
\end{eqnarray}
This single definition, depending on the location of the parameters $\tau_1,\,\,\tau_2\,\in\gamma$, generates at once four different propagators:
\begin{eqnarray}
\Delta_{--}(\mathbf{x}_1,t_1;\mathbf{x}_2,t_2) &=& -i\langle \mathcal{T} \psi(\mathbf{x}_1, t_{1-}) \psi^{\dagger}(\mathbf{x}_2,t_{2-}) \rangle \nonumber\\
&=& -i\langle \mathcal{T} \psi_{-}(\mathbf{x}_1, t_{1}) \psi_{-}^{\dagger}(\mathbf{x}_2,t_{2}) \rangle\nonumber\\
&=& -i\langle \hat{T} \psi_{-}(\mathbf{x}_1, t_{1}) \psi_{-}^{\dagger}(\mathbf{x}_2,t_{2}) \rangle\nonumber\\
&=& -i\langle \hat{T} \psi(\mathbf{x}_1, t_{1}) \psi^{\dagger}(\mathbf{x}_2,t_{2}) \rangle\;,\\
\Delta_{-+}(\mathbf{x}_1,t_1;\mathbf{x}_2,t_2) &=& -i\langle \mathcal{T} \psi(\mathbf{x}_1, t_{1-}) \psi^{\dagger}(\mathbf{x}_2,t_{2+}) \rangle \nonumber\\
&=& -i\langle \mathcal{T} \psi_{-}(\mathbf{x}_1, t_{1}) \psi_{+}^{\dagger}(\mathbf{x}_2,t_{2}) \rangle\nonumber\\
&=& -i\zeta \langle \psi_{+}^{\dagger}(\mathbf{x}_2,t_{2}) \psi_{-}(\mathbf{x}_1, t_{1})  \rangle\nonumber\\
&=& -i\zeta \langle  \psi^{\dagger}(\mathbf{x}_2,t_{2}) \psi(\mathbf{x}_1, t_{1})\rangle\;, \\
\Delta_{+-}(\mathbf{x}_1,t_1;\mathbf{x}_2,t_2) &=& -i\langle \mathcal{T} \psi(\mathbf{x}_1, t_{1+}) \psi^{\dagger}(\mathbf{x}_2,t_{2-}) \rangle \nonumber\\
&=& -i\langle \mathcal{T} \psi_{+}(\mathbf{x}_1, t_{1}) \psi_{-}^{\dagger}(\mathbf{x}_2,t_{2}) \rangle\nonumber\\
&=&  -i\langle \psi_{+}(\mathbf{x}_1, t_{1}) \psi_{-}^{\dagger}(\mathbf{x}_2,t_{2})  \rangle\nonumber\\
&=&  -i\langle  \psi(\mathbf{x}_1, t_{1}) \psi^{\dagger}(\mathbf{x}_2,t_{2}) \rangle\;,\\
\Delta_{++}(\mathbf{x}_1,t_1;\mathbf{x}_2,t_2) &=& -i\langle \mathcal{T} \psi(\mathbf{x}_1, t_{1+}) \psi^{\dagger}(\mathbf{x}_2,t_{2+}) \rangle\nonumber\\
&=& -i\langle \mathcal{T} \psi_{+}(\mathbf{x}_1, t_{1}) \psi_{+}^{\dagger}(\mathbf{x}_2,t_{2}) \rangle\nonumber\\
&=&  -i\langle \tilde{T}\psi_{+}(\mathbf{x}_1, t_{1}) \psi_{+}^{\dagger}(\mathbf{x}_2,t_{2})  \rangle\nonumber\\
&=&  -i\langle  \tilde{T} \psi(\mathbf{x}_1, t_{1}) \psi^{\dagger}(\mathbf{x}_2,t_{2})  \rangle\;.
\end{eqnarray}
Here, we have defined the usual time-order $\hat{T}$ and anti-time-order $\tilde{T}$ operators.
Notice that not all correlators are independent, since they satisfy
\begin{eqnarray}
\Delta_{+-}(x,y) + \Delta_{-+}(x,y) = \Delta_{--}(x,y) + \Delta_{++}(x,y).
\end{eqnarray}
It is customary to organize the correlators above in the matrix form
\begin{eqnarray}
\Delta(x,y) = \left[\begin{array}{cc}
\Delta_{--}(x,y) & \Delta_{-+}(x,y)\\\Delta_{+-}(x,y) & \Delta_{++}(x,y)
\end{array}
\right]\;.
\end{eqnarray}
Using the definitions above, the retarded and advances correlators can be expressed as linear combinations
of the previous ones
\begin{eqnarray}
\Delta^{A}(\mathbf{x}_1,t_1;\mathbf{x}_2,t_2) &=& i\theta(t_2 - t_1)\langle \left[ \psi(\mathbf{x}_1,t_1),\psi^{\dagger}(\mathbf{x}_2,t_2) \right]_{-\zeta}\rangle \nonumber\\
&=& \Delta_{--}(\mathbf{x}_1,t_1;\mathbf{x}_2,t_2) - \Delta_{+-}(\mathbf{x}_1,t_1;\mathbf{x}_2,t_2)\nonumber\\
&=& \Delta_{-+}(\mathbf{x}_1,t_1;\mathbf{x}_2,t_2) - \Delta_{++}(\mathbf{x}_1,t_1;\mathbf{x}_2,t_2)\;,\\
\Delta^{R}(\mathbf{x}_1,t_1;\mathbf{x}_2,t_2) &=& -i\theta(t_1 - t_2)\langle \left[ \psi(\mathbf{x}_1,t_1),\psi^{\dagger}(\mathbf{x}_2,t_2) \right]_{-\zeta}\rangle \nonumber\\
&=& -i\theta(t_1 - t_2)\Bigg( \langle \psi(\mathbf{x}_1,t_1)\psi^{\dagger}(\mathbf{x}_2,t_2)\rangle \nonumber\\
&&-\zeta \langle \psi^{\dagger}(\mathbf{x}_2,t_2)\psi(\mathbf{x}_1,t_1) \rangle\Bigg)\nonumber\\
&=& \Delta_{--}(\mathbf{x}_1,t_1;\mathbf{x}_2,t_2) - \Delta_{-+}(\mathbf{x}_1,t_1;\mathbf{x}_2,t_2)\nonumber\\
&=& \Delta_{+-}(\mathbf{x}_1,t_1;\mathbf{x}_2,t_2) - \Delta_{++}(\mathbf{x}_1,t_1;\mathbf{x}_2,t_2)\;.
\label{eq_Delta_retarded}
\end{eqnarray}
From the CTP generating functional defined in Eq.~(\ref{eq_CTP_gen}), it is possible to generate the average current components
\begin{eqnarray}
    -i  \frac{\delta\log Z_{\gamma}[A]}{\delta A_\mu(x)}&=&
    \frac{1}{Z_{\gamma}[A]}  \int \mathcal{D} \psi^\dagger \mathcal{D}\psi \, e^{\displaystyle
    i \int_{\gamma} d^3 y \mathcal{L}(y)} j^\mu(x)\nonumber\\
      &=&   \left\langle j^\mu(x) \right\rangle =:J^\mu[A](x)\,,\label{L10}
\end{eqnarray}
while the second functional derivative gives the current-current correlation function,
\begin{eqnarray}
     (-i )^2  \frac{\delta^2\log Z_{\gamma}[A]}{\delta A_\mu(x) \delta A_\nu(y)}&=&
    -i  \frac{\delta J^\mu[A](x)}{\delta A_\nu(y)}\nonumber\\
   &=&-i  \left\langle \frac{\delta j^\mu(x)}{\delta A_\nu(y)} \right\rangle +
   \left\langle \mathcal{T} j^\mu(x) j^\nu(y) \right\rangle\nonumber\\
	&&-
   \left\langle j^\mu(x)  \right\rangle \left\langle j^\nu(y) \right\rangle\,,\label{correl-js}
\end{eqnarray}
where the first term is the \emph{diamagnetic contribution} \cite{Altland-Simons}
\begin{equation}\label{diamagnetic-term}
   \begin{array}{c} \displaystyle
      \left\langle \frac{\delta j^\mu(x)}{\delta A_\nu(y)} \right\rangle=
      \delta^{\mu k} \delta^{\nu}_k \left( - \frac{{e^2}}{m^2}\right) \left\langle \psi^\dagger(x) \psi(x)\right\rangle \delta^{(3)} \left( x-y \right)
      \\ \\  \displaystyle
     =  - \frac{{e}}{m^2}\, \delta^{\mu k} \delta^{\nu}_k \left\langle j^0(x) \right\rangle \delta^{(3)}\left( x-y \right)\,,
   \end{array}
\end{equation}
and the others are the \emph{paramagnetic} ones.

We take the currents in normal order with respect to the fermionic field, so that $J^\mu[A=0]=0$. The \emph{\emph{linear response}} of the system to the external electromagnetic field is described by the second derivative in Eq.~(\ref{correl-js}) evaluated at $A_\mu=0$~\cite{Altland-Simons},
\begin{eqnarray}
      K^{\mu\nu}(x,y) &=& \left. (-i )^2  \frac{\delta^2\log Z_{\gamma}[A]}{\delta A_\mu(x) \delta A_\nu(y)} \right|_{A=0}
			= K^{\nu\mu}(y,x) \nonumber\\
     &=&\left\langle \mathcal{T} j^\mu(x) j^\nu(y) \right\rangle_0\,. \label{Kmunu}
\end{eqnarray}
Then, the density response is
\begin{eqnarray}
   K^{00}(x,y) &=& \left\langle \mathcal{T} j^0(x) j^0(y) \right\rangle_0\nonumber\\
	&=& e^2 \left\langle \mathcal{T }\psi^\dagger(x) \psi(x)\,\psi^\dagger(y) \psi(y)   \right\rangle_0\,.\label{{K00}}
\end{eqnarray}
The spatial components of the current are given by
\begin{eqnarray}
j^{k}(x)\Bigg|_{A=0} &=& \frac{e}{2m}\left\{
i\partial_{k}\psi^{\dagger}(x) \psi(x) - i\psi^{\dagger}(x) \partial_{k}\psi(x) + 2\theta \psi^{\dagger}(x)\sigma_{k}\psi(x)
\right\}\nonumber\\
&=& \psi^{\dagger}(x)\left( \frac{e}{2m}\left\{
-i\overleftrightarrow{\partial}_{k} + 2\theta \sigma_k
\right\} \right)\psi (x)\nonumber\\
&\equiv& \psi^{\dagger}_a(x)\hat{D}_{ab}^{k}\psi_b(x)\;.
\end{eqnarray}
Here, we have defined the differential operators
\begin{eqnarray}
\hat{D}_{ab}^{k} = \frac{e}{2m}\left\{-i\overleftrightarrow{\partial}_{k} \delta_{ab}+ 2\theta \left[\sigma_k\right]_{ab}
\right\}\;.
\end{eqnarray}
Applying Wick's theorem on the CTP for the definition of the current-correlator (correlators associated to disconnected diagrams vanish):
\begin{eqnarray}
 \langle \mathcal{T} j^{k}(x) j^{l} (y)\rangle
&=& \langle  \mathcal{T} \psi_{a}^{\dagger}(x)\hat{D}_{ab}^k \psi_{b}(x) \psi_c^{\dagger}(y)\hat{D}_{cd}^{l}\psi_d(y)\rangle\nonumber\\
&=& -\hat{D}_{ab}^k \hat{D}_{cd}^{l}
  \langle  \mathcal{T} \psi_{b}(x) \psi_c^{\dagger}(y)\rangle \langle  \mathcal{T} \psi_{d}(y) \psi_a^{\dagger}(x)\rangle\;.
\end{eqnarray}
The previous relation allows us to define the corresponding components of the polarization tensor in the CTP contour indices $\alpha,\beta = \pm$,
\begin{eqnarray}
K^{k l}_{\alpha\beta}(x,y) &=& \langle  \mathcal{T} j^{k}_{\alpha}(x) j^{l}_{\beta} (y)\rangle\nonumber\\
&=& -\hat{D}_{ab}^k \hat{D}_{cd}^{l} \Delta_{bc}^{\alpha\beta}(x,y)\Delta_{da}^{\beta\alpha}(y,x)\;.
\end{eqnarray}
The retarded component of the polarization tensor is obtained following the general prescription explained in Eq.(\ref{eq_Delta_retarded}),
\begin{eqnarray}
K_{R}^{kl}(x,y) &=& K_{--}^{kl}(x,y) - K_{-+}^{kl}(x,y)\nonumber\\
&=& \hat{D}_{ab}^k \hat{D}_{cd}^{l} \left\{  \Delta_{bc}^{--}(x,y)\Delta_{da}^{--}(y,x) -   \Delta_{bc}^{-+}(x,y)\Delta_{da}^{+-}(y,x)  \right\}\nonumber\\
&=& \hat{D}_{ab}^k \hat{D}_{cd}^{l} \Bigg\{
\Delta_{bc}^{F}(x,y)\Delta_{da}^{F}(y,x) \nonumber\\
&&- \left( \Delta_{bc}^{F}(x,y) - \Delta_{bc}^{R}(x,y) \right)\left( \Delta_{da}^{F}(y,x) - \Delta_{da}^{A}(y,x)  \right)
\Bigg\}\nonumber\\
&=& \hat{D}_{ab}^k \hat{D}_{cd}^{l} \Bigg\{
\Delta_{bc}^{F}(x,y) \Delta_{da}^{A}(y,x) + \Delta_{bc}^{R}(x,y) \Delta_{da}^{F}(y,x) \nonumber\\
&&- \Delta_{bc}^{R}(x,y) \Delta_{da}^{A}(y,x)
\Bigg\}\;.
\end{eqnarray}
In terms of Fourier transforms,
\begin{equation}\label{psi-de-p}
        \psi(x)= \frac{1}{\left( 2 \pi\right)^{3/2}} \int d^3p \, e^{-i p\cdot x} \tilde{\psi}(p)\,,
  \qquad
      \psi^\dagger(x)= \frac{1}{\left( 2 \pi\right)^{3/2}} \int d^3p \, e^{i p\cdot x} \tilde{\psi}^\dagger(p)\,,
    \end{equation}
we have
\begin{eqnarray}
\Delta_{ab}^{\alpha\beta}(x,y) \equiv \Delta_{ab}^{\alpha\beta}(x-y) = \int\frac{d^3 p }{(2\pi)^3} e^{i(x-y)\cdot p}\tilde{\Delta}_{ab}^{\alpha\beta}(p).
\end{eqnarray}
Here, the different propagators for the Hamiltonian model considered are, in Fourier space (F: Feynman, R: Retarded, A: Advanced),
\begin{eqnarray}
\tilde{\Delta}^{F}(p) &=& \tilde{\Delta}_{--}(p)= i\frac{\p_0 - \frac{\mathbf{p}^2}{2m} + v_F \mathbf{p}\cdot\vec{\sigma}}{\left( p_0 - \frac{\mathbf{p}^2}{2m}\right)^2 - v_F^2\mathbf{p}^2 + i\epsilon'}\nonumber\\
&&= i\frac{\p_0 - \frac{\mathbf{p}^2}{2m} + v_F \mathbf{p}\cdot\vec{\sigma}}{\left( p_0 + i\epsilon - \frac{\mathbf{p}^2}{2m} - v_F|\mathbf{p}|\right)\left( p_0 - i\epsilon - \frac{\mathbf{p}^2}{2m} + v_F|\mathbf{p}|\right)}\;,
\\
\tilde{\Delta}^{R}(p) &=& i\frac{p_0 - \frac{\mathbf{p}^2}{2m} + v_F \mathbf{p}\cdot\vec{\sigma} }{\left( p_0+ i\epsilon - \frac{\mathbf{p}^2}{2m}\right)^2 - v_F^2\mathbf{p}^2 }\;,\\
\tilde{\Delta}^{A}(p) &=& i\frac{ p_0 - \frac{\mathbf{p}^2}{2m} + v_F \mathbf{p}\cdot\vec{\sigma}}{\left( p_0 - i\epsilon - \frac{\mathbf{p}^2}{2m}\right)^2 - v_F^2\mathbf{p}^2 }\;.
\label{eq_Deltas_Fourier}
\end{eqnarray}
In particular, for the linear response theory, we need the retarded component of the polarization tensor
\begin{eqnarray}
K_{R}^{\mu\nu}(x-y) = \int\frac{d^3 p}{(2\pi)^3} e^{i(x-y)\cdot p}\, \Pi_{R}^{\mu\nu}(p)\;.
\end{eqnarray}
Here, the Fourier transform of the retarded component is given by
\begin{eqnarray}
\Pi_R^{kl}(p) = \Pi_{FA}^{kl}(p) + \Pi_{RF}^{kl}(p) - \Pi_{RA}^{kl}(p)\;,
\label{eq_PiR}
\end{eqnarray}
where the different terms are defined by
\begin{eqnarray}
\Pi^{kl}_{FA}(p) &=& \frac{e^2}{4 m^2}\int\frac{d^3 q}{(2\pi)^3}\Gamma_{ab}^k(p+2q)
\tilde{\Delta}_{bc}^{F}(p+q)
\Gamma_{cd}^l(p+2q)
\tilde{\Delta}_{da}^{A}(q)\;,\nonumber\\
\Pi^{kl}_{RF}(p) &=& \frac{e^2}{4 m^2}\int\frac{d^3 q}{(2\pi)^3}
\Gamma_{ab}^k(p+2q)
\tilde{\Delta}_{bc}^{R}(p+q)
\Gamma_{cd}^l(p+2q)
\tilde{\Delta}_{da}^{F}(q)\;,\nonumber\\
\Pi^{kl}_{RA}(p) &=& \frac{e^2}{4 m^2}\int\frac{d^3 q}{(2\pi)^3}
\Gamma_{ab}^k(p+2q)
\tilde{\Delta}_{bc}^{R}(p+q)
\Gamma_{cd}^l(p+2q)
\tilde{\Delta}_{da}^{A}(q)\;,
\label{eq_PiRAF}
\end{eqnarray}
with the symbol
\begin{equation}
\Gamma_{ab}^k(p+2q)= \left[ \delta_{ab}(p + 2q)_k + 2\theta\left[\sigma_k\right]_{ab}\right],
\end{equation}
and a similar expression for $\Gamma_{cd}^l(p+2q)$. Below we obtain the polarization tensor explicitly.

\section{The polarization tensor}\label{vacpol}

The polarization tensor $ \Pi^{kl}(p)$ contains the information about the conductivity on the plane of this two-dimensional system and also about its properties of transmission of light through it \cite{FV-2016,Altland-Simons}.
We are interested in the consequences of the application of harmonic homogeneous electric fields which, in the temporal gauge, are related with the vector potential by $E_k=-\partial A_k/\partial t=i \omega A_k$. Since the conductivity is determined by the linear relation between the current and the applied electric field, $J_k=\sigma_{kl} E_l$, from Eqs.\ \eqref{L10}, \eqref{Kmunu} and \eqref{eq_PiR}, we can write for the conductivity as a function of the frequency \cite{FV-2016,Altland-Simons}
\begin{equation}\label{sigma-pi}
    \sigma_{kl}(\omega)=\left. \frac{\Pi_{kl}^{R}(p)}{ p_0} \right|_{p \rightarrow (\omega, \mathbf{0})}\,.
\end{equation}
So, in the following we evaluate $\Pi_{kl}^{R}(\omega,\mathbf{0})$ from Eq.(\ref{eq_PiR}), that hence requires the evaluation of the three integrals defined in Eq.(\ref{eq_PiRAF}).
Let us start with $\Pi_{kl}^{FA}(p)$,
\begin{eqnarray}
\Pi_{kl}^{FA}(p) &=&\frac{e^2}{4 m^2}\nonumber\\
&&\hspace{-18mm} \times\int\frac{d^3 q}{(2\pi)^3}{\rm{Tr}}\left\{\left[ p_k + 2q_k + 2\theta \sigma_k \right]\Delta^{F}(p+q)
\left[ p_l + 2q_l + 2\theta\sigma_l\right]\Delta^{A}(q)\right\}.
\end{eqnarray}
Specializing this expression to the case $p = \left(\omega,\mathbf{0}\right)$, we write
\begin{eqnarray}
\Pi_{kl}^{FA}(\omega,\mathbf{0}) = \frac{e^2}{4 m^2}\int\frac{d^3 q}{(2\pi)^3}\frac{{\rm{Tr}}\{A\}}{B^{FA}}
\label{eq_PiFA1}
\end{eqnarray}
with
\begin{eqnarray}
A&=&\left[ 2q_k + 2\theta \sigma_k \right]\left[ \omega + q_0 - \frac{\mathbf{q}^2}{2m} + v_F \mathbf{q}\cdot\vec{\sigma}\right]
\left[ 2q_l + 2\theta\sigma_l\right]\nonumber\\
&&\times\left[ q_0 - \frac{\mathbf{q}^2}{2m} + v_F \mathbf{q}\cdot\vec{\sigma}\right]\;, \nonumber\\
B^{FA}&=&\left( \omega + q_0 + i\epsilon - \frac{\mathbf{q}^2}{2m} - v_F|\mathbf{q}|\right)\left( \omega + q_0 - i\epsilon - \frac{\mathbf{q}^2}{2m} + v_F|\mathbf{q}|\right)\nonumber\\
&&\times\left(\left( q_0 - i\epsilon - \frac{\mathbf{q}^2}{2m}\right)^2 - v_F^2\mathbf{q}^2\right)\;.\label{PiFA_den}
\end{eqnarray}
By writing $q_1=Q \cos\varphi, q_2=Q \sin\varphi$, and noticing that the denominator is independent of $\varphi$, it is straightforward to get for the trace in the numerator integrated over $\varphi$,
\begin{eqnarray}\label{trazas-integ}
     \int_0^{2\pi} {\rm Tr} \left\{ A \right\} d\varphi &=&   -8\pi\left( Q^2 + 2 m^2 v_F^2\right)q_0^2  \nonumber\\
		&&+ 8 \frac{\pi}{m} \left( Q^4 -\omega \left(m Q^2 + 2m^3 v_F^2 \right) - 2 m^2 Q^2 v_f^2\right)q_0\nonumber\\
&&     + 2 \frac{\pi}{ m^2} Q^2\left(2 m \omega \left(Q^2 -2m^2 v_F^2\right) + 2 m^2 Q^2 v_F^2 - Q^4 \right),
\end{eqnarray}
for $k,l=1,1$ or $2,2$, and a vanishing result for $k,l=1,2$ or $2,1$.

Since the previous result is a quadratic polynomial in $q_0$, and the denominator in Eq.~\eqref{PiFA_den} is a quartic expression in the integration variable, the integral over $q_0$ can be done on the complex plane, taking into account the position of the simple poles of the integrand with respect to the real axis. With a dimensional regularization of the resulting integral (dimension $d=2-s$), as described in detail in Appendix, one finds
\begin{eqnarray}
\Pi_{11}^{FA}(\omega,\mathbf{0}) = \frac{e^2}{4 m^2}\left\{ i\frac{m^2\omega}{4\pi s} - i\frac{m^2\omega}{4\pi}\log\left[ -\frac{(\omega + 2 i \epsilon)}{2 m v_F}\right]\right\}\;.
\label{eq_PiFA}
\end{eqnarray}
A similar procedure, as described in Appendix, leads to the corresponding expressions for the other two pieces of the retarded polarisation tensor
\begin{eqnarray}
\Pi_{11}^{RA}(\omega,\mathbf{0}) &=& \frac{e^2}{4 m^2}\Bigg\{ i\frac{m^2\omega}{2\pi s} - i\frac{m^2\omega}{4\pi}\log\left[ -\frac{(\omega + 2 i \epsilon)}{2 m v_F}\right]\nonumber\\
&&- i\frac{m^2\omega}{4\pi}\log\left[ \frac{(\omega + 2 i \epsilon)}{2 m v_F}\right]\Bigg\}\;,
\label{eq_PiRA} \\
\Pi_{11}^{RF}(\omega,\mathbf{0}) &=& \frac{e^2}{4 m^2}\left\{ i\frac{m^2\omega}{4\pi s} - i\frac{m^2\omega}{4\pi}\log\left[ -\frac{(\omega + 2 i \epsilon)}{2 m v_F}\right]\right\}\;.
\label{eq_PiRF}
\end{eqnarray}
We notice that the three separate parts above, which together yield the retarded polarization tensor, display a pole at $s=0$. However, when
added together according to Eq.(\ref{eq_PiR}), the poles exactly cancel to yield a finite result
\begin{eqnarray}
\Pi_{11}^{R}(\omega,\mathbf{0}) &=& \lim_{\epsilon\rightarrow 0^{+}}\Pi_{11}^{RF}(\omega,\mathbf{0}) + \Pi_{11}^{FA}(\omega,\mathbf{0}) - \Pi_{11}^{RA}(\omega,\mathbf{0})\nonumber\\
&=& \frac{e^2}{4 m^2}\frac{m^2\omega}{4} = \frac{e^2\omega}{16}\;.
\end{eqnarray}
The result above must be multiplied by a factor of 2 due to the spin degeneracy, and another factor of 2 due to valley degeneracy.  Thus, 
the final result of the optical conductivity is
\begin{eqnarray}
\sigma_{11} = 2\cdot2 \frac{\Pi_{11}^{R}(\omega,\mathbf{0})}{\omega} = \frac{e^2}{4}\;.
\end{eqnarray}
Remarkably, this is the same result that is obtained for the usual linear dispersion approximation. Therefore, we conclude that
the optical conductivity, and hence the transparency in graphene are not affected by next-to-nearest neighbor contributions
to the tight-binding microscopic model, that translate into a quadratic correction to the kinetic energy, as considered in this work.

\section{Conclusions}\label{conclusions}

Among the many outstanding properties of graphene which can be described within the Dirac limit, its optical transparency is entirely explained in terms of the fine structure constant. A natural question is to ask the extent at which such picture deviates from the experimental measurements. In this regard, in this article we considered the next-to-nearest neighbors contribution which in the continuum corrects the kinetic term with a quadratic contribution. Introducing the CTP formalism, we calculate the linear response current-current correlator from which the optical conductivity is derived.  Within this formalism, it is straightforward to obtain the retarded part of the polarization tensor after a dimensional regularization of the involved integrals. Remarkably and somehow unexpectedly, we found the conductivity of the Dirac limit to be robust against quadratic corrections. This encouraging results opens the possibility of testing deviations of the Dirac limit in graphene in other physical phenomena. These results are currently under scrutiny and results will be reported elsewhere.

\section*{Appendix: Regularization of the momentum integrals}\label{app}
In this Appendix, we present in detail the dimensional regularization method used to calculate the momentum integrals defined in the main text. Let us consider the term Eq.~(\ref{eq_PiFA1}). After taking the trace and performing the angular integral as shown in Eq.~(\ref{trazas-integ}), we have to evaluate
\begin{eqnarray}
\Pi_{11}^{FA}(\omega,\mathbf{0}) &=& \frac{e^2}{4 m^2}\int_{0}^{\infty} \frac{dQ}{(2\pi)^3} Q \int_{-\infty}^{\infty} \frac{dq_0}{B^{FA}}
\Bigg\{
-8\pi\left( Q^2 + 2 m^2 v_f^2\right)q_0^2\nonumber\\
&&+ 8\frac{\pi}{m} \left( Q^4 -\omega \left(m Q^2 + 2m^3 v_f^2 \right)
- 2 m^2 Q^2 v_f^2\right)q_0   \nonumber\\
&&+ 2 \frac{\pi}{m^2} Q^2\left(2 m \omega \left(Q^2 -2m^2 v_f^2\right) + 2 m^2 Q^2 v_f^2 - Q^4 \right)
\Bigg\}\;,
\end{eqnarray}
with $B^{FA}$ given in Eq.~(\ref{PiFA_den}) with $|\mathbf{q}|=Q$.
%
Clearly, on the $q_0$-plane, the integrand has three poles on the positive imaginary plane at $q_{0}^{(1,2)} = i\epsilon + \frac{Q^2}{2m} \pm v_F Q$, $q_{0}^{(3)} = i\epsilon-\omega + \frac{Q^2}{2m} - v_F Q$, and a single pole on the negative imaginary plane at $q_{0}^{(4)} = -i\epsilon - \omega + \frac{Q^2}{2m} + v_F Q$. We evaluate the $q_0$ integral
by means of the residue theorem, closing the contour on the lower plane. The result of this procedure can be expressed as
\begin{eqnarray}
\Pi_{11}^{FA}(\omega,\mathbf{0}) = \frac{e^2}{4 m^2}\int_{0}^{\infty} \frac{dQ}{(2\pi)^3}
\frac{Q \, P^{FA}(Q,\omega)}{\left(\omega + 2 i \epsilon\right)\left(\omega - 2 v_F Q + 2 i \epsilon\right)\left(i v_F Q + \epsilon \right)}.
\end{eqnarray}
Here, we have defined the numerator as the polynomial function
\begin{eqnarray}
P^{FA}(Q,\omega) &=& \left(-16 \pi ^2 m^2 v_F^4+16 \pi ^2 m \omega v_F^2+32 i \pi ^2 m v_F^2 \epsilon -8 i \pi ^2 \omega \epsilon +8 \pi ^2 \epsilon ^2\right)\nonumber\\
&&\hspace{-5mm}+Q \left(16 \pi ^2 m^2 \omega v_F^3+32 i \pi ^2 m^2 v_F^3 \epsilon \right)\nonumber\\
&&\hspace{-5mm}+ Q^2 \left(16 \pi ^2 m^2 v_F^2 \epsilon ^2-16 i \pi ^2 m^2 \omega v_F^2 \epsilon \right)\nonumber\\
&&\hspace{-5mm}+Q^3 \left(-32 \pi ^2 m v_F^3+8 \pi ^2 \omega v_F+16 i \pi ^2 v_F \epsilon \right)-16 Q^4 \left(\pi ^2 v_F^2\right).
\end{eqnarray}
By simply counting powers in numerator and denominator, it is clear that the remaining integral is divergent and needs regularization. For this purpose,
we first perform a partial fraction decomposition of the denominator as follows
\begin{eqnarray}
\frac{1}{\left(\omega + 2 i \epsilon\right)\left(\omega - 2 v_F Q + 2 i \epsilon\right)\left(i v_F Q + \epsilon \right)} &=&
\frac{1}{2 i\omega v_F^2\left(\omega + 2 i \epsilon\right)}\nonumber\\
&&\times\left(\frac{1}{Q - Q_1} - \frac{1}{Q - Q_2} \right)\;,\label{partial}
\end{eqnarray}
with $Q_1 = i \epsilon/v_F$, $Q_2 = \left( \omega + 2 i \epsilon\right)/(2 v_F)$. After this, the integral
splits into two contributions
\begin{eqnarray}
\Pi_{11}^{FA}(\omega,\mathbf{0}) &=& \frac{e^2}{4 m^2}\frac{1}{2 i \omega v_F^2\left(\omega + 2 i \epsilon\right)}\Bigg\{ \int_{0}^{\infty} \frac{dQ}{(2\pi)^3} Q
\frac{P^{FA}(Q,\omega)}{Q - Q_1} \nonumber\\
&&- \int_{0}^{\infty} \frac{dQ}{(2\pi)^3} Q
\frac{P^{FA}(Q,\omega)}{Q - Q_2}  \Bigg\}\;.
\end{eqnarray}
For each integral (i.e. $Q_j = Q_1, Q_2$ respectively), let us analyze the asymptotic behavior of the integrand at large momentum values, say for $Q > Q^{*}$, with
$Q^{*}$ an arbitrary but large momentum scale. In this regime,
\begin{eqnarray}
\frac{P^{FA}(Q,\omega)}{Q - Q_j} &\sim&
\frac{8 \pi ^2 Q_j}{Q^2} \Bigg(Q_j^2 \left(-2 m^2 v_F^4+2 m v_F^2 (\omega+2 i \epsilon )+\epsilon  (\epsilon -i \omega)\right)
\nonumber\\
&&+2 m^2 Q_j v_F^3 (\omega+2 i \epsilon )-2 i m^2 \omega v_F^2 \epsilon
+2 m^2 v_F^2 \epsilon ^2\nonumber\\
&&+Q_j^3 v_F \left(-4 m v_F^2+\omega+2 i \epsilon \right)-2 Q_j^4 v_F^2\Bigg)\nonumber\\
&&+8 \pi ^2 Q \Bigg(-2 m^2 v_F^4
+\omega \left(2 m v_F^2+Q_j v_F-i \epsilon \right)-4 m Q_j v_F^3\nonumber\\
&&+4 i m v_F^2 \epsilon -2 Q_j^2 v_F^2+2 i Q_j v_F \epsilon +\epsilon ^2\Bigg)\nonumber\\
&&+\frac{8 \pi ^2}{Q} \Bigg(Q_j^2 \left(-2 m^2 v_F^4+2 m v_F^2 (\omega+2 i \epsilon )+\epsilon  (\epsilon -i \omega)\right)
\nonumber\\
&&+2 m^2 Q_j v_F^3 (\omega+2 i \epsilon ) -2 i m^2 \omega v_F^2 \epsilon +2 m^2 v_F^2 \epsilon ^2\nonumber\\
&&+Q_j^3 v_F \left(-4 m v_F^2+\omega+2 i \epsilon \right)-2 Q_j^4 v_F^2\Bigg)\nonumber\\
&&+8 \pi ^2 \Bigg(Q_j \left(-2 m^2 v_F^4
+2 m v_F^2 (\omega+2 i \epsilon )+\epsilon  (\epsilon -i \omega)\right)\nonumber\\
&&+2 m^2 \omega v_F^3+4 i m^2 v_F^3 \epsilon +Q_j^2 v_F \left(-4 m v_F^2+\omega+2 i \epsilon \right)\nonumber\\
&&-2 Q_j^3 v_F^2\Bigg)
+8 \pi ^2 Q^2 v_F \left(-4 m v_F^2+\omega-2 Q_j v_F+2 i \epsilon \right)\nonumber\\
&&-16 \pi ^2 q^3 v_F^2 + O[Q^{-3}]
\nonumber\\
&\equiv &P_{FA}^{Asymp}(Q,\omega,Q_j) + O[Q^{-3}]\;,
\end{eqnarray}
where we have defined $P_{FA}^{Asymp}(Q,\omega,Q_j)$ as the polynomial obtained by truncating the asymptotic expansion above up to $O[Q^{-3}]$, for $Q > Q^{*}$.
Therefore, using this expansion, we regularize each of the integrals using the prescription ($d = 2 - s$)
\begin{eqnarray}
\int_{0}^{\infty} \frac{dQ}{(2\pi)^3} Q
\frac{P^{FA}(Q,\omega)}{Q - Q_j} &\rightarrow& \int_{0}^{Q^{*}} \frac{dQ}{(2\pi)^3} Q
\frac{P^{FA}(Q,\omega)}{Q - Q_j}\nonumber\\
& +& \int_{Q^{*}}^{\infty} \frac{dQ}{(2\pi)^3} Q
\left[\frac{P^{FA}(Q,\omega)}{Q - Q_j} - P_{FA}^{Asymp}(Q,\omega,Q_j)\right]\nonumber\\
& +& \int_{Q^{*}}^{\infty} \frac{dQ}{(2\pi)^3} m^{s}Q^{1-s}
P_{FA}^{Asymp}(Q,\omega,Q_j)\;.
\end{eqnarray}
After lengthy  but straightforward algebra, we obtain in the limit $\epsilon \rightarrow 0^{+}$
\begin{eqnarray}
\Pi_{11}^{FA}(\omega,\mathbf{0}) = \frac{e^2}{4 m^2}\left\{ i\frac{m^2\omega}{4\pi s} - i\frac{m^2\omega}{4\pi}\log\left[ -\frac{(\omega + 2 i \epsilon)}{2 m v_F}\right]\right\}\;.
\end{eqnarray}

Let us now consider the expression for $\Pi_{11}^{RF}(\omega,\mathbf{0})$, as obtained after calculating the trace and angular integration according to Eq.(\ref{trazas-integ})
\begin{eqnarray}
\Pi_{11}^{RF}(\omega,\mathbf{0}) &=& \frac{e^2}{4 m^2}\int_{0}^{\infty} \frac{dQ}{(2\pi)^3} Q \int_{-\infty}^{\infty} \frac{dq_0}{B^{RF}} \Bigg\{
-8\pi\left( Q^2 + 2 m^2 v_f^2\right)q_0^2  \nonumber\\
&&+ 8 \frac{\pi}{m} \left( Q^4 -\omega \left(m Q^2 + 2m^3 v_f^2 \right)
- 2 m^2 Q^2 v_f^2\right)q_0\nonumber\\
&& + 2 \frac{\pi}{m^2} Q^2\left(2 m \omega \left(Q^2 -2m^2 v_f^2\right) + 2 m^2 Q^2 v_f^2 - Q^4 \right)
\Bigg\}\;,
\end{eqnarray}
with
\begin{eqnarray}
B^{RF}&=&
\left(q_0 + i\epsilon - \frac{Q^2}{2m} - v_FQ\right)
\left( q_0 - i\epsilon - \frac{Q^2}{2m} + v_F Q\right)\nonumber\\
&&\times\left(\left(\omega+ q_0 + i\epsilon - \frac{Q^2}{2m}\right)^2 - v_F^2Q^2\right)\;.
\end{eqnarray}
In this case, on the $q_0$-plane the integrand has three poles on the negative imaginary plane, $q_{0}^{(1,2)} = -i\epsilon -\omega + Q^{2}/(2m) \pm v_F Q$, $q_{0}^{(3)} = -i\epsilon + Q^2/(2m) + v_F Q$,
and a single pole on the positive imaginary plane at $q_{0}^{(4)} = i\epsilon + \frac{Q^2}{2m} - v_F Q$. Therefore, we calculate the integral over $q_0$ using the residue theorem, by choosing a contour that closes
on the upper complex plane. Thus,
\begin{eqnarray}
\Pi_{11}^{RF}(\omega,\mathbf{0}) &=& \frac{e^2}{4 m^2}\nonumber\\
&&\hspace{-5mm}\times\int_{0}^{\infty} \frac{dQ}{(2\pi)^3}
\frac{Q \, P^{RF}(Q,\omega)}{\left(\omega + 2 i \epsilon\right)\left(\omega - 2 v_F Q + 2 i \epsilon\right)\left(i v_F Q + \epsilon \right)}\;.
\end{eqnarray}
The numerator of the resulting integrand is defined by the quartic polynomial function
\begin{eqnarray}
P^{RF}(Q,\omega) &=& \left(16 \pi ^2 m^2 v_F^2 \epsilon ^2-16 i \pi ^2 m^2\omega v_F^2 \epsilon \right)\nonumber\\
&&\hspace{-8mm}+Q \left(16 \pi ^2 m^2\omega v_F^3+32 i \pi ^2 m^2 v_F^3 \epsilon \right)\nonumber\\
&&\hspace{-8mm}+Q^2 (-16 \pi ^2 m^2 v_F^4-16 \pi ^2 m\omega v_F^2-32 i \pi ^2 m v_F^2 \epsilon -8 i \pi ^2\omega \epsilon 
+8 \pi ^2 \epsilon ^2)\nonumber\\
&&\hspace{-8mm}+Q^3 \left(32 \pi ^2 m v_F^3+8 \pi ^2\omega v_F+16 i \pi ^2 v_F \epsilon \right)-16 Q^4 \left(\pi ^2 v_F^2\right),
\end{eqnarray}
and then the integral is clearly divergent. A consistent regularization procedure is applied in this case as well.
By performing the same partial fraction expansion of the denominator, as in Eq.(\ref{partial}), we find that the integral splits into
two pieces ($Q_1 = i\epsilon/v_F$, $Q_2 = (\omega + 2 i\epsilon)/(2 v_F)$)
\begin{eqnarray}
\Pi_{11}^{RF}(\omega,\mathbf{0}) &=& \frac{e^2}{4 m^2}\frac{1}{2 i \omega v_F^2\left(\omega + 2 i \epsilon\right)}\Bigg\{ \int_{0}^{\infty} \frac{dQ}{(2\pi)^3} Q
\frac{P^{RF}(Q,\omega)}{Q - Q_1} \nonumber\\
&&- \int_{0}^{\infty} \frac{dQ}{(2\pi)^3} Q\frac{P^{RF}(Q,\omega)}{Q - Q_2}  \Bigg\}\;.
\end{eqnarray}
For each integral (i.e. $Q_j = Q_1, Q_2$ respectively), we analyze the asymptotic behavior of the integrand at large momentum values, say for $Q > Q^{*}$. In this regime,
\begin{eqnarray}
\frac{P^{RF}(Q,\omega)}{Q - Q_j} &&\sim
8 \pi^2 \frac{Q_j}{Q^2}\Bigg(
-2 i m^2 \omega v_F^2 \epsilon +2 m^2 v_F^2 \epsilon^2 +2 m^2Q_j v_F^3 (\omega+2 i \epsilon )
\nonumber\\
&&+Q_j^2 \left(-2 m^2 v_F^4-2 m v_F^2 (\omega+2 i \epsilon )+\epsilon  (\epsilon -i \omega)\right)\nonumber\\
&&+Q_j^3 v_F \left(4 m v_F^2+\omega+2 i \epsilon \right)-2Q_j^4 v_F^2\Bigg)\nonumber\\
&&+\frac{8 \pi ^2}{Q} \Bigg(Q_j^2 \left(-2 m^2 v_F^4-2 m v_F^2 (\omega+2 i \epsilon )+\epsilon  (\epsilon -i \omega)\right)\nonumber\\
&&+2 m^2Q_j v_F^3 (\omega+2 i \epsilon )-2 i m^2 \omega v_F^2 \epsilon +2 m^2 v_F^2 \epsilon ^2\nonumber\\
&&+Q_j^3 v_F \left(4 m v_F^2+\omega+2 i \epsilon \right)-2Q_j^4 v_F^2\Bigg)\nonumber\\
&&+8 \pi ^2 \Bigg(Q_j \left(-2 m^2 v_F^4-2 m v_F^2 (\omega+2 i \epsilon )+\epsilon  (\epsilon -i \omega)\right)+2 m^2 \omega v_F^3\nonumber\\&&+4 i m^2 v_F^3 \epsilon +Q_j^2 v_F \left(4 m v_F^2+\omega+2 i \epsilon \right)-2Q_j^3 v_F^2\Bigg)\nonumber\\
&&+8 \pi ^2 Q \Bigg(-2 m^2 v_F^4+\omega \left(-2 m v_F^2+Q_j v_F-i \epsilon \right)+4 mQ_j v_F^3\nonumber\\
&&-4 i m v_F^2 \epsilon -2Q_j^2 v_F^2+2 iQ_j v_F \epsilon +\epsilon ^2\Bigg)\nonumber\\
&&+8 \pi ^2 Q^2 v_F \left(4 m v_F^2+\omega-2Q_j v_F+2 i \epsilon \right)-16 \pi ^2 Q^3 v_F^2 + O[Q^{-3}]
\nonumber\\
&&\equiv P_{RF}^{Asymp}(Q,\omega,Q_j) + O[Q^{-3}],
\end{eqnarray}
where we have defined $P_{RF}^{Asymp}(Q,\omega,Q_j)$ as the polynomial obtained by truncating the asymptotic expansion above up to $O[Q^{-3}]$, for $Q > Q^{*}$.
Therefore, using this expansion, we regularize each of the integrals using the prescription ($d = 2 - s$)
\begin{eqnarray}
\int_{0}^{\infty} \frac{dQ}{(2\pi)^3} Q
\frac{P^{RF}(Q,\omega)}{Q - Q_j} &\rightarrow& \int_{0}^{Q^{*}} \frac{dQ}{(2\pi)^3} Q
\frac{P^{RF}(Q,\omega)}{Q - Q_j}\nonumber\\
& +& \int_{Q^{*}}^{\infty} \frac{dQ}{(2\pi)^3} Q
\left[\frac{P^{RF}(Q,\omega)}{Q - Q_j} - P_{RF}^{Asymp}(Q,\omega,Q_j)\right]\nonumber\\
& +& \int_{Q^{*}}^{\infty} \frac{dQ}{(2\pi)^3} m^{s}Q^{1-s}
P_{RF}^{Asymp}(Q,\omega,Q_j)\;.
\end{eqnarray}
After  straightforward manipulations, we obtain in the limit $\epsilon \rightarrow 0^{+}$
\begin{eqnarray}
\Pi_{11}^{RF}(\omega,\mathbf{0}) = \frac{e^2}{4 m^2}\left\{ i\frac{m^2\omega}{4\pi s} - i\frac{m^2\omega}{4\pi}\log\left[ -\frac{(\omega + 2 i \epsilon)}{2 m v_F}\right]\right\}\;.
\end{eqnarray}

Finally, let us consider the term Eq.(\ref{eq_PiRAF}). After tracing and performing the angular integral,
\begin{eqnarray}
\Pi_{11}^{RA}(\omega,\mathbf{0}) &=& \frac{e^2}{4 m^2}\int_{0}^{\infty} \frac{dQ}{(2\pi)^3} Q \int_{-\infty}^{\infty} \frac{dq_0}{B^{RA}} \Bigg\{
-8\pi\left( Q^2 + 2 m^2 v_f^2\right)q_0^2  \nonumber\\
&&+ 8\frac{\pi}{m} \left( Q^4 -\omega \left(m Q^2 + 2m^3 v_f^2 \right)
- 2 m^2 Q^2 v_f^2\right)q_0 \nonumber\\
&&  + 2 \frac{\pi}{m^2} Q^2\left(2 m \omega \left(Q^2 -2m^2 v_f^2\right) +  m^2 Q^2 v_f^2 - Q^4 \right)
\Bigg\}\;,
\end{eqnarray}
with
\begin{eqnarray}
B^{RA}&=&\left( \left(\omega + q_0 + i\epsilon - \frac{Q^2}{2m}\right)^2 - v_F^2Q^2\right)\nonumber\\
&&\times
\left( \left(q_0 - i\epsilon - \frac{Q^2}{2m}\right)^2 - v_F^2 Q^2\right)\;.
\end{eqnarray}
Clearly, on the $q_0$-plane, the integrand has two poles on the positive imaginary plane at $q_{0}^{(1,2)} = i\epsilon + \frac{Q^2}{2m} \pm v_F Q$,  and two poles on the negative imaginary plane at $q_{0}^{(3,4)} = -i\epsilon - \omega + \frac{Q^2}{2m} \pm v_F Q$. We evaluate the $q_0$ integral
by the residue theorem, closing the contour on the upper plane. Thus,
\begin{eqnarray}
\Pi_{11}^{RA}(\omega,\mathbf{0}) &=& \frac{e^2}{4 m^2}\nonumber\\
&&\hspace{-5mm}\int_{0}^{\infty} \frac{dQ}{(2\pi)^3} Q
\frac{P^{RA}(Q,\omega)}{\left(\omega + 2 i \epsilon\right)\left(\omega - 2 v_F Q + 2 i \epsilon\right)\left(\omega + 2  v_F Q + 2 i \epsilon \right)}\;.
\end{eqnarray}
The numerator of the resulting integrand is defined by the quartic polynomial function
\begin{eqnarray}
P^{RA}(Q,\omega) &=& -16 i \pi ^2 \Bigg(\omega^2 \left(2 m^2 v_F^2+Q^2\right)+2 i \omega \epsilon  \left(2 m^2 v_F^2+Q^2\right)\nonumber\\
&&-2 \epsilon ^2 \left(2 m^2 v_F^2+Q^2\right)-4 Q^2 v_F^2 \left(m^2 v_F^2+Q^2\right)\Bigg),
\end{eqnarray}
and hence the diverging integral needs also a regularization. As in the former two cases, we first do a partial fraction decomposition of the denominator,
to obtain
\begin{eqnarray}
\Pi_{11}^{RA}(\omega,\mathbf{0}) &=& \frac{e^2}{4 m^2}\frac{\left( Q_3 - Q_4 \right)^{-1}}{-4  v_F^2\left(\omega + 2 i \epsilon\right)}\Bigg\{ \int_{0}^{\infty} \frac{dQ}{(2\pi)^3} Q
\frac{P^{RA}(Q,\omega)}{Q - Q_3}\nonumber\\
&& - \int_{0}^{\infty} \frac{dQ}{(2\pi)^3} Q
\frac{P^{RA}(Q,\omega)}{Q - Q_3}  \Bigg\},
\end{eqnarray}
where we have defined $Q_3 = (\omega + 2 i \epsilon)/(2 v_F)$, $Q_4 = - Q_3$.
For each integral (i.e. $Q_j = Q_3, Q_4$ respectively), we analyze the asymptotic behavior of the integrand at large momentum values, say for $Q > Q^{*}$. In this regime,
\begin{eqnarray}
\frac{P^{RA}(Q,\omega)}{Q - Q_j} &&\sim
-16 i \pi ^2 \frac{Q_j}{Q^2} \Bigg(\omega^2 \left(2 m^2 v_F^2+Q_j^2\right)+2 i \omega \epsilon  \left(2 m^2 v_F^2+Q_j^2\right)
\nonumber\\
&&
-2 \left(Q_j^2 \left(2 m^2 v_F^4+\epsilon ^2\right)+2 m^2 v_F^2 \epsilon ^2+2 Q_j^4 v_F^2\right)\Bigg)\nonumber\\
&&-16 i \pi ^2 Q \Bigg(-2 \left(2 m^2 v_F^4+2 Q_j^2 v_F^2+\epsilon ^2\right)+\omega^2+2 i \omega \epsilon \Bigg)\nonumber\\
&&-\frac{16 i \pi ^2}{Q} \Bigg(\omega^2 \left(2 m^2 v_F^2+Q_j^2\right)+2 i \omega \epsilon  \left(2 m^2 v_F^2+Q_j^2\right)\nonumber\\
&& -2 \left(Q_j^2 \left(2 m^2 v_F^4+\epsilon ^2\right)+2 m^2 v_F^2 \epsilon ^2+2 Q_j^4 v_F^2\right)\Bigg)\nonumber\\
&&-16 i \pi ^2 Q_j \Bigg(-2 \left(2 m^2 v_F^4+2 Q_j^2 v_F^2+\epsilon ^2\right)+\omega^2+2 i \omega \epsilon \Bigg)\nonumber\\
&&+64 i \pi ^2  v_F^2 Q^3+64 i \pi ^2  Q_j v_F^2 Q^2 + O[Q^{-3}]\nonumber\\
&&\equiv P_{RA}^{Asymp}(Q,\omega,Q_j) + O[Q^{-3}],
\end{eqnarray}
where we have defined $P_{RA}^{Asymp}(Q,\omega,Q_j)$ as the polynomial obtained by truncating the asymptotic expansion above up to $O[Q^{-3}]$, for $Q > Q^{*}$.
Therefore, using this expansion, we regularize each of the integrals using the prescription ($d = 2 - s$)
\begin{eqnarray}
\int_{0}^{\infty} \frac{dQ}{(2\pi)^3} Q
\frac{P^{RA}(Q,\omega)}{Q - Q_j} &\rightarrow& \int_{0}^{Q^{*}} \frac{dQ}{(2\pi)^3} Q
\frac{P^{RA}(Q,\omega)}{Q - Q_j}\nonumber\\
&&+ \int_{Q^{*}}^{\infty} \frac{dQ}{(2\pi)^3} Q
\left[\frac{P^{RA}(Q,\omega)}{Q - Q_j} - P_{RA}^{Asymp}(Q,\omega,Q_j)\right]\nonumber\\
&& + \int_{Q^{*}}^{\infty} \frac{dQ}{(2\pi)^3} m^{s}Q^{1-s}
P_{RA}^{Asymp}(Q,\omega,Q_j)\;.
\end{eqnarray}
After lengthy but straightforward algebra, we obtain in the limit $\epsilon \rightarrow 0^{+}$
\begin{eqnarray}
\Pi_{11}^{RA}(\omega,\mathbf{0}) &=& \frac{e^2}{4 m^2}\Bigg\{ i\frac{m^2\omega}{2\pi s} - i\frac{m^2\omega}{4\pi}\log\left[ -\frac{(\omega + 2 i \epsilon)}{2 m v_F}\right]\nonumber\\
&&- i\frac{m^2\omega}{4\pi}\log\left[ \frac{(\omega + 2 i \epsilon)}{2 m v_F}\right]\Bigg\}.
\end{eqnarray}
Notice that the final result does not depend on the arbitrary scale $Q^*$, as it must.

\ack
H.F. thanks ANPCyT, CONICET and UNLP, Argentina, for partial support through grants PICT-2014-2304, PIP 2015-688 and Proy. Nro. 11/X748, respectively.
M.L. acknowledges support from FONDECYT (Chile) under grants No. 1170107, No. 1150471, No. 1150847 and ConicytPIA/BASAL (Chile) Grant No. FB0821.
E. M. acknowledges support from FONDECYT (Chile) under grant No. 1141146.
A.R.\ acknowledges support from Consejo Nacional de Ciencia y Tecnolog\'ia (Mexico) under grant  256494 and FONDECYT (Chile) under grant 1150847. H.F.\ and A.R\  also acknowledge PUC for hospitality.

\end{document}